# Characterization of the previous normal-dose CT scan induced nonlocal means regularization method for low-dose CT image reconstruction

Hao Zhang, Jianhua Ma, William Moore, and Zhengrong Liang*

*Abstract*—Repeated computed tomography (CT) scans are required in some clinical applications such as image-guided radiotherapy and follow-up observations over a time period. To optimize the radiation dose utility, a normal-dose (or full-dose) CT scan is often first performed to set up reference, followed by a series of low-dose scans. Using the previous normal-dose scan to improve follow-up low-dose scans reconstruction has drawn great interests recently, such as the previous normal-dose induced nonlocal means (ndiNLM) regularization method. However, one major concern with this method is that whether it would introduce false structures or miss true structures when the previous normal-dose image and current low-dose image have different structures (e.g., a tumor could be present, grow, shrink or absent in either image). This study aims to investigate the performance of the ndiNLM regularization method in the above mentioned situations. A patient with lung nodule for biopsy was recruited to this study. A normal-dose scan was acquired to set up biopsy operation, followed by a few low-dose scans during needle intervention toward the nodule. We used different slices to mimic different possible cases wherein the previous normal-dose image and current low-dose image have different structures. The experimental results characterize performance of our ndiNLM regularization method.

## I. INTRODUCTION

X-ray computed tomography (CT) has been widely exploited in clinic for different applications. Recent discoveries regarding the potential harmful effects of X-ray radiation including genetic and cancerous diseases have raised growing concerns to patients and medical physics community [1]. Repeated CT scans are required in some clinical applications such as image-guided radiotherapy and follow-up observations over a time period, and the accumulated radiation dose could be significant. To optimize radiation dose utility, a normal-dose scan is often first performed to set up reference, followed by a series of low-dose scans. In these applications, the previous normal-dose scan can be exploited as prior information due to the similarity among the reconstructed image series of the scans. While somewhat misalignment and/or deformation may occur among the image series, they can be mitigated through registration of the image series. Using the reconstruction from previous normal-dose scan to improve the follow-up low-dose scans reconstruction has become a research interest recently, some of which exploited the previous normal-dose image as a penalty for regularized iterative image reconstructions. For instance, Nett et al. [2] incorporated a registered normal-dose image into their prior image constrained compressed sensing (PICCS) framework for iterative reconstruction of subsequent low-dose CT images. Stayman et al. [3, 4] presented a PICCS-type penalty term, but the high-quality normal-dose image was formulated into a joint estimation framework for both image registration and image reconstruction in order to better capturing the anatomical motion among different scans. Zhang et al. [5, 6] predicted MRF coefficients from previous normal-dose CT image and exploited this prior information to improve the follow-up statistical Bayesian low-dose image reconstruction. Moreover, Ma et al. [7, 8] proposed previous normal-dose image induced nonlocal means (ndiNLM) penalty terms to improve the following low-dose CT image reconstruction for perfusion and interventional imaging, wherein the previous normal-dose scan was also pre-registered with the low-dose scans. Because of the patch-based search mechanism, this approach does not heavily depend on the accuracy of registration, and a rough registration would be adequate in practice [7, 8]. However, one major concern with the ndiNLM regularization method is that whether it would introduce false structures or miss true structures when the previous normal-dose image and current low-dose image have different structures (e.g., a tumor could be present, grow, shrink or absent in either the previous or current image). This study investigates the different scenarios wherein the previous normal-dose image and current low-dose image have different structures, and characterizes the performance of the ndiNLM regularization method in these situations.

## II. METHODS

### A. Statistical model

The noise property of the calibrated line integrals has been investigated by analyzing experimental data of a physical phantom from repeated scans. The statistical analysis showed that the calibrated line integrals can be fitted approximately by a Gaussian distribution with a nonlinear signal-dependent variance [9, 10]:

$$y_i \sim Gaussian(\bar{y}_i, \sigma_{y_i}^2) \qquad (1)$$

This work was supported in part by NIH/NCI under grants #CA143111 and #CA082402.
H. Zhang is with the Departments of Radiology and Biomedical Engineering, State University of New York at Stony Brook, NY 11794, USA.
J. Ma is with the School of Biomedical Engineering, Southern Medical University, Guangdong 510515, China.
W. Moore is with the Department of Radiology, State University of New York at Stony Brook, NY 11794, USA.
*Z. Liang is with the Departments of Radiology and Biomedical Engineering, State University of New York at Stony Brook, NY 11794, USA (jerome.liang@sunysb.edu).

With the 'Poisson+Gaussian' noise model for the transmitted photons, it has been shown in that the variance of the line integral $y_i$ can be given by [11, 12]:

$$\sigma_{y_i}^2 = \frac{\bar{N}_i + \sigma_e^2}{\bar{N}_i^2} = \frac{1}{\bar{N}_{0i}}\exp(\bar{y}_i)\left(1 + \frac{\sigma_e^2}{\bar{N}_{0i}}\exp(\bar{y}_i)\right) \quad (2)$$

where $\bar{N}_{0i}$ represents the mean number of X-ray photons just before entering the patient and going toward the detector bin $i$, and $\sigma_e^2$ denotes the variance of the electronic noise.

### B. PWLS image reconstruction

The penalized weighted least-squares (PWLS) cost function in the image domain can be written as:

$$\Phi(\boldsymbol{\mu}) = (\mathbf{y} - \mathbf{A}\boldsymbol{\mu})^T \boldsymbol{\Sigma}^{-1}(\mathbf{y} - \mathbf{A}\boldsymbol{\mu}) + \beta R(\boldsymbol{\mu}) \quad (3)$$

where $\mathbf{y} = (y_1,...,y_I)^T$ is the vector of measured line integrals, and $I$ is the number of projection measurements; $\boldsymbol{\mu} = (\mu_1,...,\mu_J)^T$ is the vector of attenuation coefficients of the object to be reconstructed, and $J$ is the number of image pixels; $\mathbf{A}$ is the projection matrix with the size $I \times J$, and its element $A_{ij}$ is typically calculated as the intersection length of projection ray $i$ with pixel $j$. In our implementation, the system matrix is pre-calculated by a fast ray-tracing technique and stored as a file to serve as a lookup table during iterations. $\boldsymbol{\Sigma}$ is the covariance matrix, and since the measurement among different detector bins are assumed to be independent, the matrix is diagonal and $\boldsymbol{\Sigma} = \text{diag}\{\sigma_{y_i}^2\}$. The symbols $T$ and $-1$ herein are transpose and inverse operators, respectively. $R(\boldsymbol{\mu})$ denotes the penalty term and $\beta$ is a smoothing parameter which plays a role of controlling the tradeoff between the data fidelity term and the penalty term. Eq. (3) is the well-known PWLS criterion in the image domain.

The goal for CT image reconstruction is to estimate the attenuation coefficients $\boldsymbol{\mu}$ from the noisy measurement $\mathbf{y}$:

$$\hat{\boldsymbol{\mu}} = \arg\min_{\mu \geq 0} \Phi(\boldsymbol{\mu}) \quad (4)$$

Minimization of Eq. (4) could also be efficiently achieved with the Gauss-Seidel update strategy, and the details can be found in a previous paper [8].

### C. ndiNLM regularization

The ndiNLM penalty in the image domain can be described as [7]:

$$R(\boldsymbol{\mu}) = \sum_j (\mu_j - \sum_{k \in SW_j} w_{jk} \mu_k^{ND})^2 \quad (5)$$

where $\boldsymbol{\mu}^{ND} = (\mu_1^{ND},...,\mu_J^{ND})^T$ denotes the vector of attenuation coefficients for the previous normal-dose image (ND is short for normal-dose), SW denotes a search-window, and the NLM-based weighting coefficients $w_{jk}$ are given as:

$$w_{jk} = \frac{\exp\left(-\|PW(\mu_j) - PW(\mu_k^{ND})\|_{2,a}^2 / h^2\right)}{\sum_{k \in SW_j}\left[\exp\left(-\|PW(\mu_j) - PW(\mu_k^{ND})\|_{2,a}^2 / h^2\right)\right]} \quad (6)$$

where PW denotes a patch-window, and h is the filtering parameter.

### III. RESULTS

#### A. Data acquisition

To evaluate the ndiNLM regularization method in a more realistic situation, a patient with lung nodule for biopsy at Stony Brook University Hospital was recruited to this study under informed consent after approval by the Institutional Review Board. The patient was scanned using a Siemens CT scanner. The X-ray tube voltage was set to be 120 kV, and the tube current was set to be 100 mAs. The raw data was calibrated by the CT system and outputted as sinogram data or line integrals. We regarded this acquisition as the previous normal-dose scan, and simulated the corresponding low-dose sinogram data by adding noise to the normal-dose sinogram data using the simulation method in [13]. The noisy measurement $N_i$ at detector bin $i$ was generated according to the statistical model:

$$N_i \sim \text{Poisson}(\bar{N}_{0i}\exp(-\bar{y}_i)) + \text{Gaussian}(0, \sigma_e^2) \quad (7)$$

Then the corresponding noisy line integral $\{y_i\}$ is calculated by the logarithm transform.

#### B. Scenario 1: low-dose CT image having nodule

Fig. 1 illustrates one transverse image of the patient from the simulated low-dose sinogram data, reconstructed by the FBP method. The red arrow indicates a lung nodule. Fig. 2 (P1)-(P3) illustrates three transverse images of the patient from the acquired normal-dose sinogram data, reconstructed by the FBP method. They serve as the previous normal-dose image for the PWLS-ndiNLM method, to mimic the different cases wherein the previous normal-dose image and current low-dose image may have different structures. Fig. 2 (R1)-(R3) show the corresponding reconstructed images from the simulated low-dose sinogram data, by the PWLS-ndiNLM method. It should be noted that the $\beta=1\times10^5$, SW=$34\times34$, PW=$5\times5$, and h=0.005 for all of our implementations. We can see that Fig. 2 (R2) and (R3) still retain the lung nodule even though the corresponding previous normal-dose image has shrinking nodule (P2) or no nodule (P3).

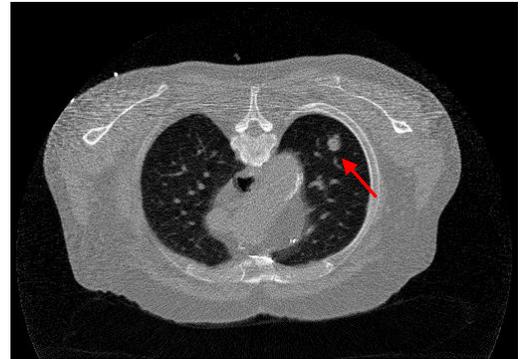

FIG. 1. One transverse image of the patient with a lung nodule, reconstructed by the FBP method from simulated low-dose sinogram data.

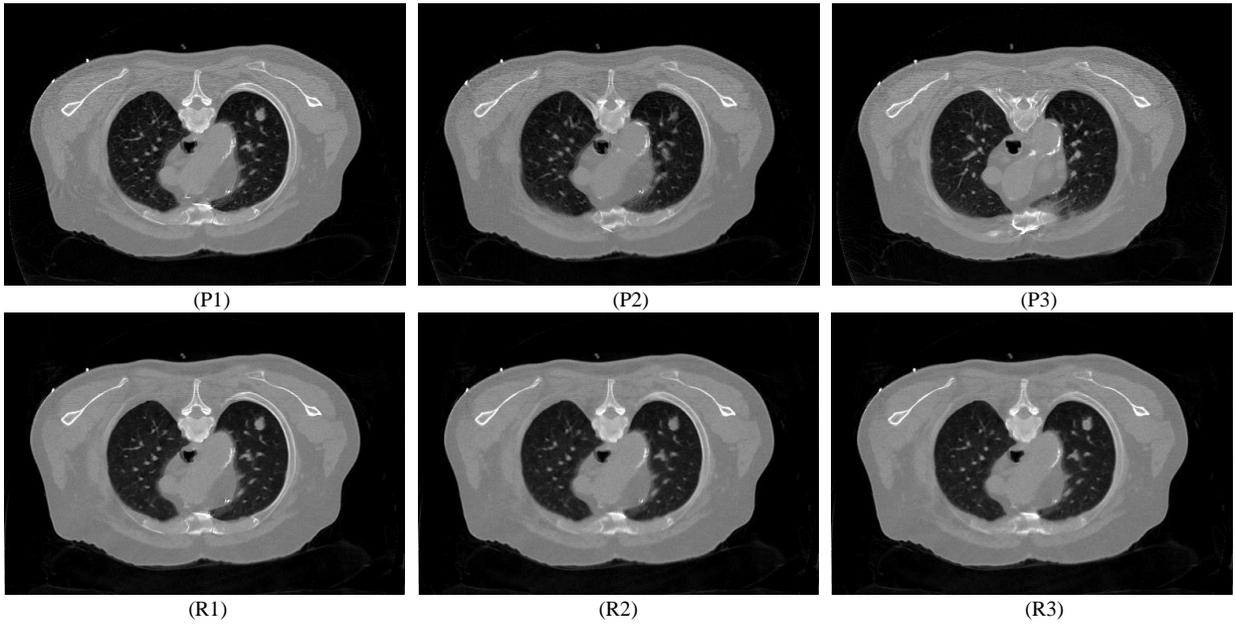

FIG.2. Transverse images of the patient. (P1)-(P3) -- three transverse images of the patient from the normal-dose sinogram data, reconstructed by the FBP method; (R1)-(R3) -- one transverse image of the patient reconstructed by the PWLS-ndiNLM method, from the simulated low-dose sinogram. All the images are displayed with the same window.

*C. Scenario 2: low-dose CT image having no nodule*

Fig. 3 illustrates another transverse image of the patient from the simulated low-dose sinogram data, reconstructed by the FBP method. It can be observed that this image has no lung nodule. Again, Fig. 4 (P1)-(P3) illustrate three transverse images of the patient from the acquired normal-dose sinogram data, reconstructed by the FBP method. They serve as the previous normal-dose image for the PWLS-ndiNLM method, to mimic the different cases wherein the previous normal-dose image and current low-dose image may have different structures. Fig. 4 (R1)-(R3) shows the corresponding reconstructed images from the simulated low-dose sinogram data, by the PWLS-ndiNLM method. We can observe that Fig. 4 (R1) and (R2) does not introduce false lung nodule when the corresponding previous normal-dose image has lung nodule.

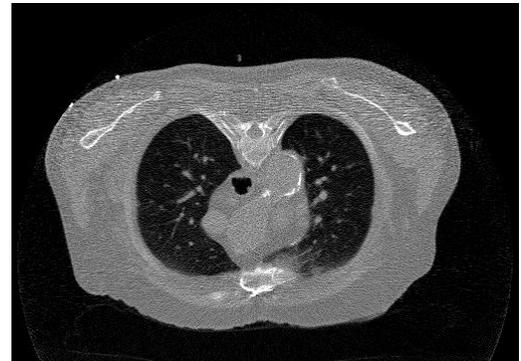

FIG. 3. Another transverse image of the patient without lung nodule, reconstructed by the FBP method from simulated low-dose sinogram data.

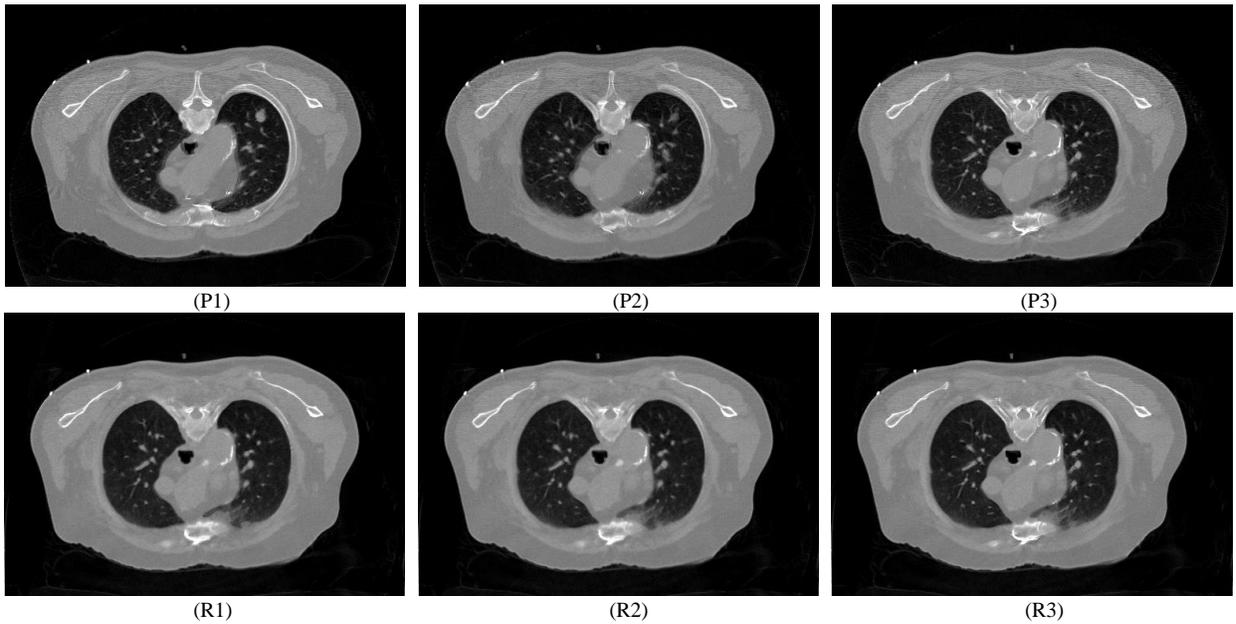

FIG.4. Transverse images of the patient. (P1)-(P3) -- three transverse images of the patient from the normal-dose sinogram data, reconstructed by the FBP method; (R1)-(R3) -- another transverse image of the patient reconstructed by the PWLS-ndiNLM method, from the simulated low-dose sinogram. All the images are displayed with the same window.

## IV. Discussions and Conclusions

In this work, we investigated the performance of the ndiNLM regularization when the previous normal-dose image and current low-dose image have different structures, for example, the normal-dose image has lung nodule but the low-dose image has no nodule, or *vice versa*. This preliminary study relieves the concern of introducing false information when the previous scan has or does not have small abnormalities into the current scan by demonstrating that the ndiNLM regularization does not introduce false nodule or miss true nodule. This is an important characteristic for the ndiNLM regularization. However, further quantitative evaluations may be needed to illustrate the reconstruction quality of whole image and lung nodule, and is currently under progress.

Compared with the generic NLM regularization [14, 15] which only utilizes current low-dose image, the ndiNLM regularization may need a larger search-window to take into account the structure difference between the previous normal-dose image and current low-dose image. Therefore, a 17×17 search-window was used in [14, 15] for the generic NLM regularization, but a 34×34 search-window was used in this study for the ndiNLM regularization. Otherwise, the ndiNLM regularization may have an inferior performance and generate undesirable results. This is another finding for this study.